%% file: bilocal-arXiv.tex
\documentclass[doublecol]{epl2}

\usepackage[active]{srcltx}
\usepackage{epsfig}
\usepackage{amsthm,mathrsfs}
\usepackage{amssymb,bm,mathtools}
\usepackage{hyperref}

\usepackage{color}
\usepackage{framed}
\definecolor{shadecolor}{rgb}{.8,.8,.8}

\input{myqcircuit}
\newcommand\Lfm{\textsc{lfm}}

\newcommand\Locc{\textsc{locc}}

\newcommand\Mes{\textsc{mes}}

\newcommand\Qt{\textsc{qt}}

\newcommand\Rqt{\textsc{rqt}}

\newcommand\Fqt{\textsc{fqt}}

\def\Reals{\mathbb{R}}

\def\Complexes{\mathbb C}

\def\spc#1{\mathcal{#1}}
\def\Hilb{\spc H}

\def\Tr{\operatorname{Tr}}

\def\Herm{\operatorname{Herm}}

\def\>{\rangle}
\def\Ket#1{|#1\>}

\def\<{\langle}
\def\Bra#1{\<#1|}

\def\KetBra#1#2{|#1\rangle\langle#2|}

\def\Projector#1{\KetBra{#1}{#1}}

\newcommand\SetStates{\mathrm{St}}

\newcommand\SetEffects{\mathrm{Eff}}

\newcommand{\sys}[1]{\mathrm{#1}}

\newcommand{\sA}{\sys{A}} 
\newcommand{\sB}{\sys{B}}
\newcommand{\sC}{\sys{C}}
\newcommand{\sM}{\sys{M}}
\newcommand{\sN}{\sys{N}}
\newcommand{\sF}{\sys{F}}

\def\ie{{i.e.~}}

\def\eg{{e.g.~}}

\begin{document}

\title{Fermionic computation is non-local tomographic and violates monogamy of entanglement}


\author{
  G.~M.~D'Ariano\inst{1,2}\thanks{E-mail: \email{dariano@unipv.it}} \and 
  F.~Manessi\inst{1}\thanks{E-mail: \email{franco.manessi01@ateneopv.it}} \and 
  P.~Perinotti\inst{1,2} \thanks{E-mail: \email{paolo.perinotti@unipv.it}} \and
  A.~Tosini\inst{1,2}\thanks{E-mail: \email{alessandro.tosini@unipv.it}} }
\shortauthor{
  G.~M.~D'Ariano \etal }
\institute{
  \inst{1} QUIT group,  Dipartimento di Fisica, via Bassi   6, 27100 Pavia, Italy. \\
  \inst{2} INFN Sezione di Pavia, via Bassi, 6, 27100 Pavia, Italy. 
}

\abstract{We show that the computational model based on local Fermionic modes in place of qubits does not
satisfy local tomography and monogamy of entanglement, and has mixed states with maximal entanglement of
formation. These features directly follow from the parity superselection rule. We generalize quantum
superselection rules to general probabilistic theories as sets of linear constraints on the convex set of
states. We then provide a link between the cardinality of the superselection rule and the degree of holism of
the resulting theory.}

\maketitle
In his pioneering paper on physical computation
\cite{feynman1982simulating}, Feynman wondered about the possibility
of simulating Fermions by local commuting quantum systems in
interaction---what we would call nowadays a ``quantum computer''. Ever
since, the relation between Fermions and local quantum systems has
been largely investigated. The Jordan-Wigner map
\cite{jordan1928paulische} transforms isomorphically the Fermionic
algebra into a qubit algebra, and has been a valuable instrument for
solving the 1d \textsc{xy} spin-chains
\cite{onsager1944crystal,lieb1964two}, or to extend to the Fermionic
case notions as the entanglement \cite{PhysRevA.76.022311}, the
entropic area law \cite{PhysRevLett.96.010404}, and universal
computation \cite{Bravyi2002210,PhysRevA.81.050303}.  However, in many
applications one needs to map quantum algebras in an ``isolocal'' way,
namely mapping local quantum operations into local ones, and nonlocal
to nonlocal ones. The Jordan-Wigner transform is not isolocal,
and this leads to some ambiguities in defining the partial trace
\cite{PhysRevA.83.062323,PhysRevA.85.016301,PhysRevA.85.016302,PhysRevA.87.022338},
and in assessing the local nature of quantum operations
\cite{verstraete2005mapping}. Here the Wigner superselection rule
comes to help.

The Wigner superselection rule forbids superpositions between states with odd and even particle
number, based on the simple argument of the impossibility of discriminating a $2\pi$ rotation from
the identity \cite{streater1964pct,weinberg1996quantum}. The Wigner superselection rule allows one
to circumvent the problems connected to isolocality \cite{PhysRevA.76.022311} without restoring it.
The price to pay, as we will see in this letter, is that the theory becomes non {\em
  locally-tomographic} \cite{PhilosophyQuantumInformationEntanglement2,chiribella2011informational},
namely one cannot discriminate between two nonlocal states using only local measurements, unlike
quantum theory (\Qt).

The notion of local-tomography (also called local-discriminability
\cite{PhysRevA.81.062348}) has been introduced in the new context of
general probabilistic theories, which has become the stage for the
recent axiomatization program of \Qt. Examples of such theories are
the classical information theory \cite{chiribella2011informational},
the box-world \cite{PhysRevA.75.032304}, and the real quantum theory
(\Rqt) \cite{stueckelberg1961quantum,hardy2012limited}. In such a framework, a
theory that lacks local-tomography is called {\em holistic}
\cite{hardy2012limited}.
In this letter we will introduce a notion of superselection rule for a general probabilistic theory,
corresponding to a linear constraint over the convex set of states. Such a notion contains the usual
superselection rules of \Qt\ as special cases, but also includes other cases, \eg \Rqt\ as a
superselection restriction. We will provide a link between the number of linearly-independent
constraints and the degree of holism of the resulting theory.

In addition to the feature of local-tomography, another characteristic trait of \Qt\ is the {\em
  monogamy of entanglement}, \ie loosely speaking a limitation on the sharing of entanglement. For
example, if two qubits are maximally entangled, neither of them can be entangled with any other
system. After extending the usual notions of \emph{entanglement of formation} and \emph{concurrence} to the
Fermionic scenario, we will show that in Fermionic quantum theory (\Fqt) entanglement is in general not
monogamous, due to the Wigner superselection rule. As we will show, the monogamy violation goes hand
in hand with the existence of maximally entangled states that are mixed. Moreover, one has
Maximally-Entangled Sets (\Mes) \cite{kraus} containing more than one bipartite state, whereas \Qt\
has only the singlet, and non-trivial \Mes's needs tripartite systems.

In the following we will restrict to probabilistic theories that are convex (\ie all sets of
transformations are convex) and causal \cite{chiribella2011informational} (namely, the probability
of the preparation is independent of the choice of the observation test).  Transformations include
as special cases states and effects, and we will denote by $\SetStates(\sA) $ and $\SetEffects(\sA)
$ the convex set of (generally subnormalized) states and the convex set of effects of system $\sA $.
In the presence of the {\em non restriction hypothesis} in its extended version (namely all
admissible transformations belong to the theory) a theory is fully specified by the sets of states
$\SetStates(\sA) $ and by composition of systems.  Imposing a superselection rule $\sigma $ on a
theory corresponds to sectioning linearly all sets of transformations for each multipartite system.
Under the non restriction hypothesis this reduces to sectioning linearly just the sets of states.
Thus, superselecting the system $ \sA $ with the rule $ \sigma $ means sectioning linearly $
\SetStates(\sA) $ giving a new set of states $ \SetStates(\bar \sA) $, which is identified with the
system $ \bar\sA \coloneqq \sigma(\sA) $ of the constrained theory.
%
%
%
For consistency, the superselection map  $\sigma $ must commute with
system composition, forcing the definition of composition for the
constrained theory as  $\sigma(\sA)\sigma(\sB)\coloneqq\sigma(\sA\sB) $
(we remind that system composition is denoted by juxtaposition, namely
the composed system of  $\sA $ and  $\sB $ is  $\sA\sB $).  Notice that,
being linear  $\sigma $ preserves convexity of the theory. This means
that \eg in a \Qt\ with a superselection rule states from different
sectors cannot be superimposed, but can be mixed.

In the following we will denote by X $_{\Reals} $ the linear span of the set X, \eg
$\SetEffects_\Reals(\sA) $ is just the set of linear functionals on states. The superselection rule
$ \sigma $ will be defined for an arbitrary system $\sA $ through linearly independent effects
$s_i^\sigma\in\SetEffects_{\Reals}(\sA) $, $i=1,\ldots V^\sigma_\sA $ as follows
\begin{equation}\label{eq:constraints}\nonumber
\SetStates[\sigma(\sA)]\coloneqq\{\rho\in\SetStates(\sA)|s_i^\sigma(\rho)= 
0, \,i=1,\ldots,
V^\sigma_\sA\}. 
\end{equation}
Clearly  $\SetStates(\bar\sA)\subseteq\SetStates(\sA) $ and
 $\SetEffects(\bar\sA)\subseteq\SetEffects(\sA) $.  One has
\begin{equation}\label{DDV}
D_{\bar\sA} = D_{\sA} - V_{\sA}^\sigma,
\end{equation}
where $ D_{\sA}\coloneqq\dim[\SetStates_{\Reals}(\sA)]  $.  For a
general theory one has $ D_{\sA\sB} \geq D_{\sA} D_{\sB}  $, and this
provides an upper bound for the number of independent constraints of a
composite system, \ie $ V_{\sA\sB}^{\sigma} \leq
D_{\sA}V_{\sB}^{\sigma} + D_{\sB}V_{\sA}^{\sigma} + D_{\sA\sB} -
V_{\sA}^{\sigma}V_{\sB}^{\sigma}  $.
 
It is easy to see that for any $b\in\SetEffects(\bar\sB) $ and any $i=1,\ldots, V^\sigma_\sA $, the
functional $ s_i^\sigma\otimes b\in \SetEffects_{\Reals}(\sA\sB) $ is a constraint for
$\overline{\sA\sB} $. Indeed, suppose by contradiction that $ s_i^\sigma\otimes b(\rho) \ne 0 $ for
$\rho\in\SetStates(\overline{\sA\sB}) $, then since $b(\rho)=\alpha $ is a valid state for $ \bar\sA
$, we have $s_i^\sigma(\alpha)\ne 0 $ against the hypothesis.  The same argument holds exchanging
the subsystems $\sA $ and $\sB $, and we conclude that the composite system $ \overline{\sA\sB} $
has at least $ D_{\bar\sA} V_{\sB}^{\sigma} + D_{\bar\sB} V_{\sA}^{\sigma} = D_{\sA}V_{\sB}^{\sigma}
+ D_{\sB}V_{\sA}^{\sigma} - 2 V_{\sA}^{\sigma}V_{\sB}^{\sigma} $ of linearly independent
constraints. In summary we have the bounds
\begin{equation}\label{eq:constraints-on-the-constraints}
\begin{aligned}
&V_{\sA\sB}^{\sigma}\geq D_{\sA}V_{\sB}^{\sigma} + D_{\sB}V_{\sA}^{\sigma} - 
2 V_{\sA}^{\sigma}V_{\sB}^{\sigma},\\
&V_{\sA\sB}^{\sigma}\leq D_{\sA}V_{\sB}^{\sigma} + D_{\sB}V_{\sA}^{\sigma}  -
V_{\sA}^{\sigma}V_{\sB}^{\sigma}+D_{\sA\sB} - D_{\sA}D_{\sB}.  
\end{aligned}
\end{equation}
A superselected theory saturating the lower bound in
Eq.~\eqref{eq:constraints-on-the-constraints} is called {\em minimal}.
For such a theory the constraints for bipartite systems are only those
of the form  $s_i^\sigma\otimes b $ and  $a\otimes s_j^\sigma $, with
 $a\in\SetEffects(\bar\sA) $ and  $b\in\SetEffects(\bar\sB) $. A minimal
superselected theory can be built ``bottom-up'' by defining the
constraints on the elementary systems.

Before proceeding, we recall the notions of  $n $-local effect and of
{\em $n $-local-tomographic} theory \cite{hardy2012limited}. We call an effect $n $-local if it can
be written as conical combination of composite effects made of effects that are at most $n$-partite.  A set of effects
E is called {\em separating} for a set of states S if any two states of S are discriminated by an
effect of E. We call a theory $n $-local-tomographic if the set of $n $-local effects is
separating for multipartite states. For a local-tomographic theory (\ie $n=1 $) one has
$D_{\sA\sB}=D_{\sA}D_{\sB} $. Notice that an $n $-local-tomographic theory is also $(n+1)
$-local-tomographic.  We will call a theory {\em strictly $n $-local-tomographic} if it is $n
$-local-tomographic but not $(n-1) $-local-tomographic.

By definition, a strictly bilocal-tomographic theory (\ie  $n=2 $) has
\cite{hardy2012limited}
\begin{align}
& \! \! \! D_{\sA\sB}>D_{\sA} D_{\sB},\label{eq:genuinely-bilocal-a}\\
& \!\! \!  D_{\sA\sB\sC}\leq D_{\sA} D_{\sB}D_{\sC}+ 
\widetilde D_{\sA \sB} D_{\sC}+\widetilde D_{\sB\sC}
D_{\sA}+\widetilde D_{\sC \sA}D_{\sB},\label{eq:genuinely-bilocal-b}
\end{align}
where 
\begin{equation}\label{DtildaAB}
\widetilde D_{\sA\sB}: = D_{\sA\sB} -D_\sA D_\sB.
\end{equation}
A strictly bilocal-tomographic theory that saturates the upper bound will be called {\em maximally
  bilocal-tomographic}, and it requires all $2 $-local effects to separate multipartite states. 


In the following we will focus on the superselection of a
local-tomographic theory. This is the case, for example, of \Qt\
with parity or charge-superselection and of the \Rqt. In this case $
D_{\sA\sB} = D_{\sA}D_{\sB}  $ and therefore
Eq.~\eqref{eq:constraints-on-the-constraints} becomes
\begin{align}
&V_{\sA\sB}^{\sigma}\geq D_{\sA}V_{\sB}^{\sigma} + D_{\sB}V_{\sA}^{\sigma} - 
2 V_{\sA}^{\sigma}V_{\sB}^{\sigma},
\label{eq:constraints-on-the-constraints-quantum-a}\\
&V_{\sA\sB}^{\sigma} \leq D_{\sA}V_{\sB}^{\sigma} + D_{\sB}V_{\sA}^{\sigma} - 
V_{\sA}^{\sigma}V_{\sB}^{\sigma}.
\label{eq:constraints-on-the-constraints-quantum-b}
\end{align}
\normalsize
%
In this scenario we have a striking relation between the discriminability of 
states and
superselection rules. Indeed a minimal superselected theory is maximally 
bilocal-tomographic.
This can be proved by evaluating $ D_{\overline{\sA\sB\sC}}  $ using
the saturated bound of
Eq.~\eqref{eq:constraints-on-the-constraints-quantum-a} and the
identities of Eq.~\eqref{DtildaAB},
Eq.~\eqref{DDV}, and $\overline{\sA\sB\sC}=\bar\sA(\overline{\sB\sC}) $, and noticing that it is
equal to the RHS of Eq.~\eqref{eq:genuinely-bilocal-b}.  While a minimal superselected theory is
maximally bilocal-tomographic, a theory saturating the upper bound
~\eqref{eq:constraints-on-the-constraints-quantum-b} is local-tomographic. In the intermediate range
one can find superselected theories of any degree of holism.

We now give a precise definition of the \Fqt. The \Fqt\ is the theory
with no-restriction hypothesis, whose generic system $\sN_\sF$ is the
quantum one made only of a finite number $N$ of qubits, satisfying the
constraint that all states must commute with the parity operator
$P_N:=\frac{1}{2}(I+\prod_{j=1}^N\sigma_z^{(j)})$ ($\sigma_\alpha $
denotes a Pauli matrix). The system $1_{\rm F}$ corresponds to the
so-called {\em local Fermionic mode} (\Lfm), and the state spaces
introduced here are the same as in Bravyi and Kitaev
\cite{Bravyi2002210}, however, the no-restriction hypothesis allows
for more transformations in our case. Locality of transformations,
necessary for defining the composition of systems, is given in terms
of the local Fermionic algebra, which is generated by the creation and
annihilation operators $\psi_i^\dag $ and $\psi_i $, respectively,
with $i=1,\ldots N$, satisfying the anti-commutation relations
$[\psi^\dag_i,\psi_j]_+=\delta_{ij}I $, $ [\psi_i,\psi_j]_+=0 $.  Let
$\ket{\Omega}$ be the unique joint eigenvector of the operators
$\psi^\dag_j\psi_j$ with zero eigenvalue, and build a representation
of field operators for a given ordering $1,2, \ldots N$ of the qubits,
given by the orthonormal basis for ${\mathbb C}^{2^N}$:
\begin{equation}\label{eq:fock-basis}
|q_1,\ldots,q_N\>:=\psi_1^{\dag q_1}
  \ldots\psi_N^{\dag q_N}|\Omega\>\quad q_i=0,1;\ i=1,\ldots, N;
\end{equation}
 where $\Ket{q_1,\ldots,q_N}$ are the joint
eigenvectors for the qubit $\sigma_z^{(j)}$ for $ j=1,\ldots, N$, forming a basis for the 
Fock space. Notice that a vector of Eq.~\eqref{eq:fock-basis}  
corresponds to a \emph{Slater
determinant} in
the first quantization formalism. The
chosen ordering identifies a specific Jordan-Wigner transform. We now
can define locality of transformations.  We say that an admissimble
transformation of the $\sN_{\rm F}$ system is local on the subsystem
$\sM_{\rm F}$ with $M<N$ if the Kraus operators belong the
representation of the field algebra of $\sM_{\rm F}$. The parity
super-selection rule forbids superpositions of vectors belonging to
$\Hilb_0$ and $\Hilb_1$ eigen-spaces of the parity operators in
${\mathbb C}^{2^ N}$, and splits the operator spaces representing
$\SetEffects_{\mathbb
  R}(\sN_{\mathrm{F}})=\SetStates_{\Reals}(\sN_{\mathrm{F}}) $ as
$\Herm(\Hilb_0)\oplus \Herm(\Hilb_1)$, whose operators are spanned by
products of even numbers of field operators.

In the following we will denote by $\sN_{\mathrm{Q}} $ the multipartite system of $N $ qubits, with
$2^N $-dimensional Hilbert space. Since \Qt\ is local-tomographic we have
$D_{\sN_{\mathrm{Q}}}=D_{1_{\mathrm{Q}}}^N= 2^{2N} $. On the other hand, according to the parity
prescription the dimension of the Fermionic system $\sN_{\mathrm{F}} $ is $
D_{\sN_{\mathrm{F}}}=2^{2N-1}=D_{\sN_{\mathrm{Q}}}/2 $. Notice that the single-\Lfm\ system
$1_{\mathrm{F}} $ has only two possible pure states $\Ket{0},\Ket{1} $, thus corresponding to the
classical bit, whereas the linear space of states for the system of $N $ \Lfm's is
$\SetStates_{\Reals}(\sN_{\mathrm{F}})=\SetEffects_{\Reals}(\sN_{\mathrm{F}})=
\Herm(\Complexes^{2^{N-1}})\oplus\Herm(\Complexes^{2^{N-1}}) $, namely the direct sum of two copies
of the state-space of $N-1 $ qubits.

The \Fqt\ saturates the bound of Eq.~\eqref{eq:genuinely-bilocal-b},
and is then maximally bilocal-tomographic.  Indeed, for
elementary Fermionic systems we have
\begin{align}\label{eq:bilocal}
\!\!\!  8=D_{2_{\mathrm{F}}}>D_{1_{\mathrm{F}}}^{2}=4,\quad
& D_{3_{\rm F}}=D_{1_{\mathrm{F}}}^3+3\widetilde
  D_{2_{\mathrm{F}}}D_{1_{\mathrm{F}}}=32
\end{align}
where $\widetilde D_{2_{\mathrm{F}}}=D_{2_{\mathrm{F}}}-D_{1_{\mathrm{F}}}^2=4 $ is the dimension of
the non-local component of $2_{\mathrm{F}} $.  The full theory is maximally bilocal-tomographic,
indeed, the number of independent local and 2-local effects for $N $ \Lfm's is $ \sum_{k=0}^{\lfloor
  n/2\rfloor} \binom {n} {2k}D_{1_{\mathrm{F}}}^{n-2k}\widetilde
D_{2_{\mathrm{F}}}^k=2^{2n-1}=D_{\sN_{\mathrm{F}}} $. We emphasize that the \Fqt\ provides an
example of a bilocal-tomographic theory whose systems do not satisfy the dimensional prescription in
Ref.~\cite{hardy2012limited} (see note \footnote{In Ref.~\cite{hardy2012limited} the
  authors after proving that $D_{\sA\sB} -D_{\sA} D_{\sB}=L_\sA L_\sB $ for some integers $L_\sA$
  and $L_\sB $, under the assumption that $ D_{\sA}+L_{\sA} $, $ D_{\sA}-L_{\sA} $ are strictly
  increasing functions of the number of perfectly discriminable states $d_{\sA} $, they prove that
  in a bilocal-tomographic theory one has $ D_\sA=\frac{1}{2}(d_{\sA}^r+d_{\sA}^s) $ for some
  integers $r,s $ satisfying $r \geq s>0 $. The strict monotonicity of the function $D_{\sA}-L_{\sA}
  $ is too restrictive and it excludes the \Fqt\ from the set of admissible bilocal-tomographic
  theories.  Indeed, for the \Fqt\ we have $D_{\sN_{\mathrm{F}}}-L_{\sN_{\mathrm{F}}}=0 $ for any
  $\sN_{\mathrm{F}} $.}).

Besides being bilocal-tomographic, the \Fqt\ is also a minimally superselected \Qt\ of qubits. It is
easy to see that the $1_{\mathrm{F}} $ system can be achieved from the qubit by means of the
superselection constraints $ \Tr[\sigma_x \rho]=\Tr[\sigma_y \rho] =0 $ for all $
\rho\in\SetStates(1_{\mathrm{F}}) $, hence
$D_{1_{\mathrm{F}}} = D_{1_{\mathrm{Q}}}-V_{1_{\mathrm{Q}}}^{\sigma} $ with
$V_{1_{\mathrm{Q}}}^{\sigma}=2 $.  The whole \Fqt\ can be built bottom-up by minimally extending the
constraints to the composite systems. Indeed the lower bound in
Eq.~\eqref{eq:constraints-on-the-constraints-quantum-b} is achieved.

Since the \Fqt\ is minimally superselected from a local-tomographic theory, it must be maximally
bilocal-tomographic. This is indeed the case, as one can see from the dimensional analysis in
Eq.~\eqref{eq:bilocal}.

It is worth mentioning that the \Fqt\ is not the only minimal
superselected \Qt.  Another example is given by \Rqt. Its systems
 $\sN_{\mathrm{R}} $ have dimensions $
D_{\sN_{\mathrm{R}}}=d_{\sN_{\mathrm{R}}}(d_{\sN_{\mathrm{R}}}+1) / 2
 $ with  $d_{\sN_{\mathrm{R}}} $ the number of perfectly distinguishable
states for the system  $\sN_{\mathrm{R}} $. On the other hand one has $
\sN_{\mathrm{R}} =\sigma(\sN_{\mathrm{Q}})  $ where the superselection
rule is given by the constraint  $\rho-\rho^T=0 $, with  $T $ denoting
transposition with respect to a fixed basis taken as real, that for
 $1_{\mathrm{R}} $ (one {\em rebit}) corresponds to the linear
constraint  $\Tr[\sigma_y\rho]=0 $. The \Rqt\ is minimally
superselected, since the number of constraints for the composite
system $ \sN_{\mathrm{R}}\sM_R  $ given by $
V_{\sN_{\mathrm{R}}\sM_R}^{\sigma}=\tfrac{1}{2}
d_{\sN_{\mathrm{R}}}d_{\sM_R}(d_{\sN_{\mathrm{R}}}d_{\sM_R}-1)  $
saturates the lower bound
\eqref{eq:constraints-on-the-constraints-quantum-b}. Then the theory
is maximally bilocal-tomographic, as pointed out in
\cite{hardy2012limited}.

Notice that, due to the parity constraint, the \Fqt\ retains only superpositions of pure states with
total occupation numbers that are equal modulo 2. If instead we allow only superpositions with total
occupation numbers that are equal modulo $k$ for any integer $k$, we get a theory that is
$k$-local-tomographic.

We now study entanglement in the \Fqt, and show that it shares some features with the \Rqt, as the
existence of maximally-entangled mixed states, and the violation of entanglement monogamy. We will
see that these phenomena are due to the fact that both theories are superselected versions of \Qt.
One would conjecture that both features may be related to the non local-tomographic nature of the
theories, however, this remains an open issue.

In a general probabilistic theory entanglement must be quantified in operational terms, namely as a
\emph{resource} for performing a task. For example, entanglement in \Qt\ represents the resource needed to
prepare states of the theory under the restriction of \Locc\ (local operations and classical
communication).  For bipartite states in \Qt\ all measures of entanglement refer to a standard unit---the {\em
ebit}---which is the amount of entanglement of a bipartite singlet state, and the so-called {\em entanglement
of formation} is the number of ebits that are needed to achieve the state by \Locc. Since the \Fqt\ is
non-classical, considering entanglement as a resource under \Locc\ is meaningful.  A full theory of
entanglement for the \Fqt\ would require a complete analysis of the transformations of states under \Locc:
this is beyond the scope of the present letter.  However, here we will show that, independently of such
analysis, one can assess features that are very different from those of entanglement in \Qt.  These are: 1)
the existence of mixed states with maximal entanglement of formation; 2) the need of \Mes\ \cite{kraus} for
bipartite states; 3) bipartite states with maximal entanglement of formation that do not belong to a \Mes; 4)
the violation of monogamy of entanglement. 

We now extend the notion of {\em concurrence} \cite{hill1997entanglement} and provide a lower bound
to {\em entanglement of formation} \cite{bennett1996mixed} for the \Fqt.

In the usual quantum scenario, the entanglement of formation is defined for a generally mixed state
$\rho\in\SetStates(\sA\sB) $ as follows $E(\rho) \coloneqq \min_{\mathcal{D}_\rho} \sum_i p_i
S(\Tr_\sA\Ket{\Psi_i}\Bra{\Psi_i}) $, where $S(\sigma)$ is the von Neumann entropy of the state
$\sigma$, and ${\mathcal D}_\rho\coloneqq \{\{ p_i, \Ket{\Psi_i} \} \mid \rho = \sum_i p_i
\Projector{\Psi_i} \} $ is the set of all the pure decomposition of the mixed state $ \rho $. One
has $E(\rho)=\lim_{n\to\infty} F_n(\rho)/n $, where $F_n(\rho) $ is the minimum number of singlets
states needed by two parties in order to prepare via \Locc\ $n$ random states $\Ket{\Psi_i}$ in any
decomposition that achieves $E(\rho)$ \cite{wootters2001entanglement}.  The bound is achieved for
pure states \cite{PhysRevA.56.R3319}.  For a mixed state $ \rho $ of two qubits one has
$E(\rho)=\mathcal{E}(C(\rho))$, with $ \mathcal{E}(x) \coloneqq h(\tfrac{1+\sqrt{1-x^2}}{2}) $, $ h
$ the binary Shannon entropy, and the concurrence $ C(\rho) $ defined as
\begin{equation}\label{eq:concurrence}
C(\rho) \coloneqq \min_{\mathcal{D}_\rho} \sum_i p_i C(\Ket{\Psi_i}),
\end{equation}
with $C(\Ket{\Psi})$ for pure states given in Ref.~\cite{hill1997entanglement}.  Both the
entanglement of formation and the concurrence are zero if and only if the state $ \rho $ is
separable, and for two qubits they reach the maximum value $1 $ if and only if $ \rho $ is a
maximally entangled state.
%
%
%

In Ref.~\cite{caves}, both the entanglement of formation and the concurrence have been specialized
to \Rqt\ restricting the minimum to the set of pure
decompositions  $\mathcal{D}_\rho^{\mathrm{R}} $ on real states. 
In Ref.~\cite{PhysRevA.76.022311} the entanglement of formation has been extended to the \Fqt; here we do the
same for the concurrence
%
\begin{align}
E_{\mathrm{F}} (\rho) &\coloneqq \min_{\mathcal{D}_\rho^{\mathrm{F}}} \sum_i p_i 
E(\Ket{\Psi_i}), \\ 
C_{\mathrm{F}} (\rho) &\coloneqq \min_{\mathcal{D}_\rho^{\mathrm{F}}} \sum_i p_i
C(\Ket{\Psi_i}),
\end{align}
with $ \mathcal{D}_\rho^{\mathrm{F}}  $  the set of all the pure
decompositions of $ \rho  $ that satisfy the parity superselection
rule. Since each mixed state is parity-decomposed uniquely as  $\rho =
p_0 \rho_0 + p_1 \rho_1 $ and all Fermionic decompositions in $
\mathcal{D}_\rho^{\mathrm{F}}  $ must preserve  $p_0 $ and  $p_1 $, one has $
E_{\mathrm{F}}(\rho) = p_0 E_{\mathrm{F}}(\rho_0) + p_1 E_{\mathrm{F}}(\rho_1)  $ and $
C_{\mathrm{F}}(\rho) = p_0 C_{\mathrm{F}}(\rho_0) + p_1 C_{\mathrm{F}}(\rho_1)  $.
Moreover, since $ \mathcal{D}_{\rho_i}^{\mathrm{F}} \equiv \mathcal{D}_{\rho_i}  $, 
we have $ E_{\mathrm{F}} (\rho_i) = E(\rho_i)  $ and $ C_{\mathrm{F}} (\rho_i) = C(\rho_i)  $, hence
\begin{align}\label{eq:sbrosbi}
E_{\mathrm{F}}(\rho) & = p_0 E(\rho_0) + p_1 E(\rho_1), \\
C_{\mathrm{F}}(\rho) & = p_0 C(\rho_0) + p_1 C(\rho_1).
\end{align}
The above definition of entanglement of formation is not proved to have the same operational
asymptotic interpretation as in \Qt, however, one can prove that it is a lower bound for it, since
bipartite fermionic \Locc's are all admissible quantum \Locc's, and any fermionic entangled resource-state  has a quantum entanglement of formation smaller than (or equal to) one.  Notice that,
unlike \Qt\ \cite{hill1997entanglement} and \Rqt\ \cite{caves}, the quantities $ E_{\mathrm{F}} $
and $ C_{\mathrm{F}} $ do not satisfy the relation $ E_{\mathrm{F}}(\rho) =
\mathcal{E}(C_{\mathrm{F}}(\rho)) $. Nevertheless we have that $ E_{\mathrm{F}}(\rho) \geq
\mathcal{E}(C_{\mathrm{F}}(\rho)) $, and for a maximally-entangled state $ \Phi $ it is $
E_{\mathrm{F}}(\Phi) = \mathcal{E}(C_{\mathrm{F}}(\Phi)) = 1 $. Therefore, when $E_F(\rho)=1$, $E_F
$ coincides with the operational entanglement of formation.
Moreover, notice that the states of Eq.\eqref{eq:fock-basis} have Fermionic entanglement of formation equal
to zero, according to the fact that a single Slater determinant in the Fermionic theory is actually a product state.

Using the quantities $ E_{\mathrm{F}}  $ and $C_{\mathrm{F}}  $ we can show 
that in the \Fqt\ there
exist maximally-entangled mixed states.  The state
\begin{equation}\label{eq:entangled-mixed}
\Phi
\coloneqq\tfrac{1}{4}\left(I\otimes I+\sigma_x\otimes\sigma_x\right),
\end{equation}
is the equal mixture of the Fermionic pure states
 $\Ket{\Psi_0}=\tfrac{1}{\sqrt{2}}\left(\Ket{00}+\Ket{11}\right) $ and
 $\Ket{\Psi_1}=\tfrac{1}{\sqrt{2}}\left(\Ket{01}+\Ket{10}\right) $.  It is easy to check that $
E_{\mathrm{F}}(\Phi) = C_{\mathrm{F}}(\Phi) = 1  $, \ie $ \Phi  $ has maximal entanglement of
formation. On the other hand in \Qt\  $\Phi  $ is separable since it can be 
regarded as
the equal mixture of the pure states $ \Ket{+} \Ket{+}  $, $ \Ket{-}\Ket{-}  $, with
 $\Ket{\pm}=\tfrac{1}{\sqrt{2}}\left(\Ket{0}\pm\Ket{1}\right) $, which gives $ E(\Phi) = C(\Phi) = 0
 $.  Such a decomposition, however, is not allowed in the Fermionic case, because the states $
\Ket{\pm}  $ violate the parity superselection rule. We could have replaced  $\sigma_x $ in Eq.
\eqref{eq:entangled-mixed} with any linear combination of  $\sigma_x $ and  $\sigma_y $ according to the
superselection constraints  $\Tr[\sigma_x\rho]=\Tr[\sigma_y\rho]=0 $ on the single \Lfm\ system. Since
in \Rqt\ we have only the linear constraint  $\Tr[\sigma_y\rho]=0 $ for one 
rebit, the
same argument holds for the state in Eq.~\eqref{eq:entangled-mixed} with 
 $\sigma_x $ replaced by
 $\sigma_y $ \cite{caves}, namely the theory has mixed maximally-entangled states.

As already mentioned, the state $ \Phi  $ despite having maximum entanglement of formation, cannot
be transformed by \Locc\ into a maximally-entangled pure state. It actually happens that for two
\Lfm's the concept of maximally-entangled state under \Locc\ has to be superseded by the concept of
\Mes, as it has already been pointed out for  $n $-partite quantum entanglement with  $n\geq3 $
\cite{kraus}. A \Mes\ for an  $n $-partite system is the minimal set of  $n $-partite states such that
any other  $n $-partite state can be obtained by \Locc\ from a state in the set. Two examples of \Mes\
for two \Lfm's are the set of all even non-factorized pure states with positive coefficients,
and the set of all odd non factorized pure states (notice that \Locc\ can change the parity using
the map $ \sigma_x \cdot\sigma_x $).

Consider now the $3_{\mathrm{F}} $ pure state $ \Ket{\Phi_p} \coloneqq \tfrac{1}{2}( \Ket{000} +
\Ket{110} + \Ket{011} + \Ket{101} ) $. If we trace the state $ \Ket{\Phi_p} $ over any one of the
three \Lfm\ we find that the reduced bipartite state of two \Lfm's is the mixed state $ \Phi $ of
Eq.~\eqref{eq:entangled-mixed} which has maximal entanglement of formation. Therefore, in the \Fqt,
as well as in \Rqt\ \cite{wootters2012monogamy}, monogamy of entanglement is violated, since the
amount of entanglement can be totally shared by each pair of systems, a feature forbidden in \Qt.

We conclude this letter by observing that, while Fermionic computation and standard quantum
computation have been shown to be equivalent \cite{Bravyi2002210}, our findings about Fermionic
entanglement suggests that the same may not hold for distributed Fermionic computation.

\acknowledgments

This work has been supported in part by the Templeton Foundation under the project ID\# 43796 {\em A
  Quantum-Digital Universe}. 


\end{document}

%% file: myqcircuit.tex
\usepackage[matrix,frame,arrow]{xy}
\usepackage{amsmath}

\newcommand{\ket}[1]{\left\vert{#1}\right\rangle}